\documentclass[12pt]{article}
\usepackage{rotate}
\usepackage{epsfig}
\usepackage{psfrag}
\usepackage{a4}

\begin{document}

\newcommand \be  {\begin{equation}}
\newcommand \bea {\begin{eqnarray} \nonumber }
\newcommand \ee  {\end{equation}}
\newcommand \eea {\end{eqnarray}}

\title{{\bf The leverage effect in financial markets: retarded volatility 
and market panic}}

\author{Jean-Philippe Bouchaud$^{1,2}$, Andrew Matacz$^2$ and Marc 
Potters$^2$}

\date{\small
$^1$ Service de Physique de l'\'{E}tat Condens\'{e}, Centre d'\'{e}tudes
de Saclay\\
Orme des Merisiers, 91191 Gif--sur--Yvette Cedex, France \\
\vspace{0.4cm}
$^{2}$ Science \& Finance\\
The research division of Capital Fund Management\\
109--111 rue Victor-Hugo, 92532 Levallois Cedex, France \\
\vspace{0.4cm}
\today
} 
\maketitle


\begin{abstract}
We investigate quantitatively the so-called leverage effect, which
corresponds to a negative correlation between past returns and
future volatility. For individual stocks, this correlation is moderate and
decays exponentially over $50$ days, while for stock indices, it is
much stronger but decays faster. For individual stocks, the magnitude
of this correlation has a universal value that can be rationalized in terms 
of a 
new `retarded' model which interpolates between a purely additive and a purely
multiplicative stochastic process. For stock indices a specific market panic 
phenomenon 
seems to be necessary to account for the observed amplitude of the effect.
\end{abstract}
\section{Introduction}

Several `stylized fact' of financial markets, such as `fat tails' in
the distribution of returns or long ranged volatility correlations,
have recently become the focus of detailed empirical study
\cite{volfluct1,volfluct2,lo,stanley,PCB,MS,Book,muzy1,muzy2}.  Simple agent 
based 
models
have been proposed, with some degree of success, to explain these
features \cite{Farmer,Bak,Lux,Hommes,MG1,MG2,BGM}. Another well-known 
stylized fact 
is the so-called `leverage'
effect, first discussed by Black \cite{Black,Cox}, who observed that
the volatility of stocks tends to increase when the price
drops. This effect is particularly important for option markets: not
only does it imply that at-the-money volatilities tend to increase
after price drops, but also that a significant skew in the volatility
smile should appear \cite{skewsmile,Book,us,fouque}, as is indeed
observed on markets where the leverage effect is strong. This skew
reflects the fact that a negative volatility-return correlation
induces a negative skew in the distribution of price returns
themselves \cite{skewsmile,us}.

Although widely discussed in the economic and econometric literature
\cite{lo,leverage}, the leverage effect (or volatility-return
correlation) has been less systematically investigated than the
volatility clustering effect (volatility-volatility correlation). For
example, one would like to know if the volatility-return correlation
shows a long term dependence similar to that observed on the
volatility-volatility correlation. Although various single correlation
coefficients quantifying the leverage effect have been measured and discussed
within {\sc garch} models \cite{lo,leverage}, 
the full temporal structure of this correlation has not been quantitatively
investigated.  The economic interpretation of
this leverage effect is furthermore still controversial; a recent
survey of the different models can be found in \cite{leverage}. Even the
causality of the effect is debated \cite{leverage}: is the volatility 
increase induced by
the price drop or conversely do prices tend to drop after a volatility 
increase?

In the present paper, we report some empirical study of this leverage effect
both for individual stocks and for stock indices. We find unambiguously that
correlations are between future volatilities and past price changes. For both 
stocks and
stock indices, the volatility-return correlation is short ranged, with
however a different decay time for stock indices (around 10 days) and
for individual stocks (around 50 days). The amplitude of the
correlation is also different, and is much stronger for indices than
for stocks. We then argue that the leverage effect for stocks can be
interpreted within a simple retarded model, where the absolute
amplitude of price changes does not follow the price level
instantaneously (as is assumed in most models of price changes, such
as the geometric Brownian motion). Rather, absolute price changes are
related to an average level of the past price. This reflects the lag
with which market operators change their behavior (order volumes,
bid-ask spreads, transaction costs, etc.) when the price evolves: the 
proportionality
between absolute price changes and the level of the price itself is
only ensured on the long run. We then show that this model does not
represent adequately the leverage effect for stock indices, which
seems to reflect a `panic'-like effect, whereby large price drops of
the market as a whole triggers a significant increase of activity.

\section{Empirical results}

\begin{figure}
\psfrag{yaxis}[ct][ct]{\small ${\cal L}(\tau)$}
\psfrag{xaxis}[cb][cb]{\small $\tau$}
\psfrag{legend1}[l][l]{\small Empirical data}
\psfrag{legend2}[l][l]{\small Exponential fit}
\centerline{\epsfig{file=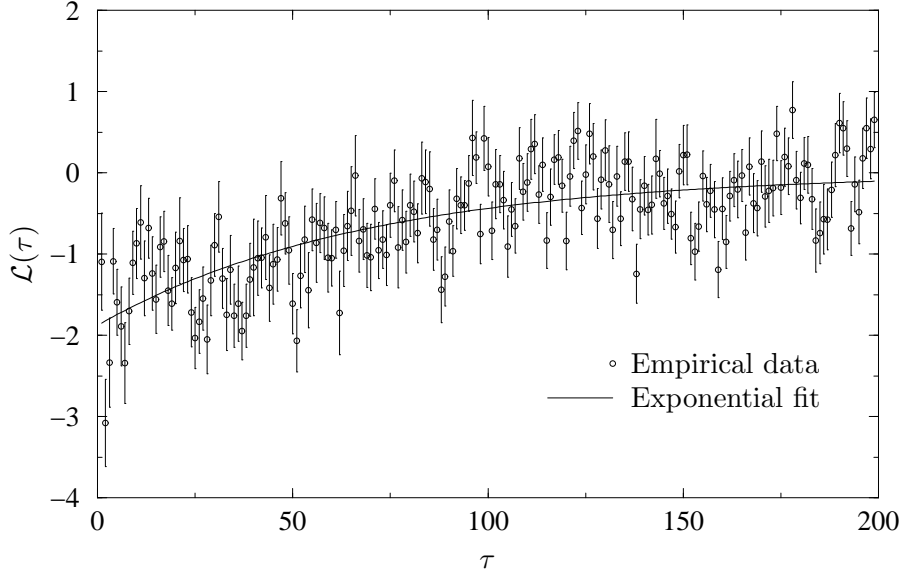,width=0.825\textwidth}
}
\caption{Return-Volatility correlation for individual
stocks. Data points are the empirical correlation averaged over 437 US
stocks, the error bars are two sigma errors bars estimated from the
inter-stock variability. The full line shows an exponential fit
(Eq.~(\protect\ref{exp_fit})) with $A_S=1.9$ and $T_S=69$ days.
Note that ${\cal L}(\tau=0)$ is not far from $-2$, as our retarded
model predicts.
}
\label{fig:fig_stocks}
\end{figure}

We will call $S_i(t)$ the price of stock $i$ at time $t$, and $\delta S_i(t)$ 
the (absolute) 
daily price change: $\delta S_i(t)=S_i(t+1)-S_i(t)$. The relative price 
change will
be denoted as $\delta x_i(t)=\delta S_i(t)/S_i(t)$. The leverage correlation 
function 
which naturally appears in the calculation of the skewness of the 
distribution of price changes over 
a horizon $T$ is \cite{us}:
\be
{\cal L}_{i}(\tau)=\frac{1}{Z} \left\langle \left[\delta x_i(t+\tau)\right]^2 
\delta x_i(t) \right\rangle,\label{C1}
\ee
which measures the correlation between price change at time $t$ and a measure 
of the
square volatility at time $t+\tau$. The coefficient $Z$ is a normalization 
that we have chosen to be:
\be
Z= \left\langle \delta x_i(t)^2 \right\rangle^2
\ee
for reasons that will become clear below. (Note that since $\delta x$ is 
dimensionless,
${\cal L}_{i}(\tau)$ is also dimensionless, despite this apparent lack of 
homogeneity between the 
numerator and denominator.)

In the following, we will consider $\tau \geq 1$. Negative values of $\tau$ 
lead to very small values of 
the correlation function, indistinguishable from noise. In other words, 
the correlation exists between future volatilities and past price changes; 
conversely, volatility changes do not convey any useful information on future 
price changes.

\begin{figure}
\psfrag{yaxis}[ct][ct]{\small ${\cal L}(\tau)$}
\psfrag{xaxis}[cb][cb]{\small $\tau$}
\psfrag{legend1}[l][l]{\small Empirical data}
\psfrag{legend2}[l][l]{\small Exponential fit}
\centerline{\epsfig{file=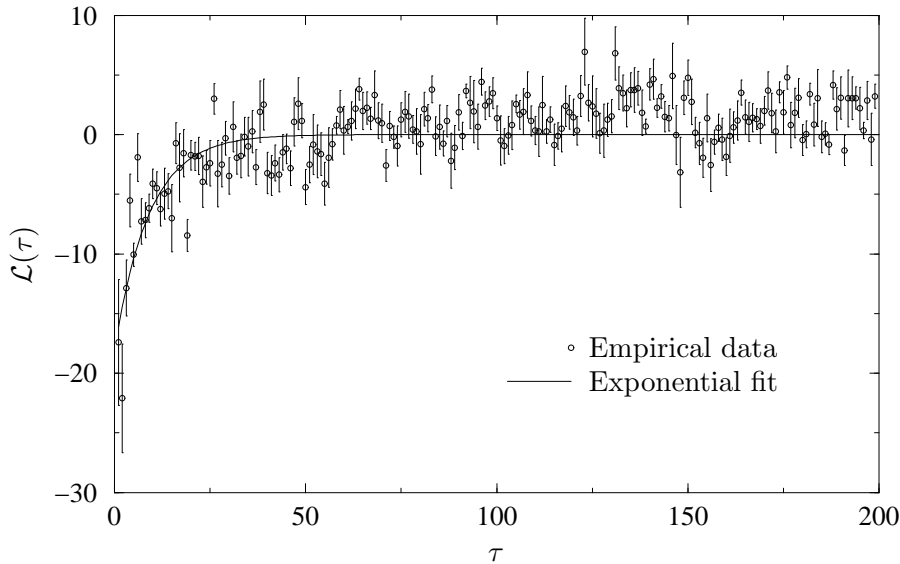,width=0.825\textwidth}}
\caption{Return-Volatility correlation for
stock indices. Data points are the empirical correlation averaged over 7
major stock indices, the error bars are two sigma errors bars estimated from 
the
inter-index variability. The full line shows an exponential fit
(Eq.~(\protect\ref{exp_fit})) with $A_I=18$ and $T_I=9.3$ days.
}
\label{fig:fig_indices}
\end{figure}

We have analyzed a set of 437 US stocks, constituent of the S\&P 500
index and a set of 7 major international stock indices. Our dataset
consisted of daily data ranging from Jan.\ 1990 to May 2000 for
stocks and from Jan.\ 1990 to Oct.\ 2000 for indices.
We have computed ${\cal L}_i$ both for individual stocks and stock indices. 
The raw results were
rather noisy. We have therefore assumed that individual stocks
behave similarly and averaged ${\cal L}_{i}$ over the 437 different stocks in 
our dataset 
to give $\overline{{\cal L}_S}$, and over 7 different indices (from the US, 
Europe and Asia), to give 
$\overline{{\cal L}_I}$. The results are given in Fig.~1 and Fig.~2 
respectively. As can be seen from these figures, both $\overline{{\cal L}_S}$ 
and
$\overline{{\cal L}_I}$ are clearly negative: price drops increase the 
volatility -- this is the 
so-called leverage effect. These correlation functions can 
rather well be fitted by {\it single} exponentials:
\be\label{exp_fit}
\overline{{\cal L}_{S,I}}(\tau)=-A_{S,I} 
\exp\left(-\frac{\tau}{T_{S,I}}\right).
\ee
For US stocks, we find $A_S=1.9$ and $T_S \simeq 69$ days, whereas for 
indices the amplitude 
$A_I$ is significantly larger, $A_I=18$ and the decay time shorter: $T_I 
\simeq 9$ days. 
This exponential decay should be contrasted with the very slow, power-law 
like decay of 
the volatility correlation function, which cannot be characterized by a 
unique decay time
\cite{volfluct2,stanley,PCB,MS,Book,muzy1,muzy2}.
Therefore, a new time scale seems to be present in financial markets, 
intermediate between 
the very high frequency time scale 
seen on the correlation function of returns (several minutes) and the very 
low frequency 
time scales appearing in the volatility correlation function.

\section{A retarded model}

Traditional models of asset fluctuations postulate that price changes are 
proportional to 
prices themselves. The price increment is therefore written as:
\be
\delta S_i(t)= S_i(t) \, \sigma_i \epsilon(t),\label{basic}
\ee
where $\sigma_i$ is the volatility and $\epsilon$ a random variable with zero 
mean 
and unit variance,
independent of all past history \cite{f1}.
Eq.~(\ref{basic}) means that price increments are at any time proportional to 
the 
current value of the price. Although it is true that on the long run, price 
increments tend
to be proportional to prices themselves, this is not reasonable on short time 
scales. Locally,
prices evolve in discrete steps (ticks), following buy or sell orders that 
can only be
expressed an integer number of contracts. The mechanisms leading
to price changes are therefore not expected to vary continuously as prices 
evolve, but rather 
to adapt only progressively if prices are seen to rise (or decrease) 
significantly 
over a certain time window. The model 
we propose to describe this lagged response to price changes is to replace 
$S_i$ in 
Eq.~(\ref{basic}) by a moving average $S_i^R$ over a certain time window. 
More precisely, we
write:
\be
\delta S_i(t) = S_i^R(t) \, \sigma_i \epsilon, \qquad 
S_i^R(t)=\sum_{\tau=0}^\infty {\cal 
K}(\tau) S_i(t-\tau),\label{retarded}
\ee
where ${\cal K}(\tau)$ is a certain averaging kernel, normalized to one:
\be
\sum_{\tau=0}^\infty {\cal K}(\tau) \equiv 1.
\ee
For example, an exponential moving average corresponds to ${\cal 
K}(\tau)=(1-\alpha) 
\alpha^\tau$, ($\alpha < 1$). It will be more congenial to rewrite $S_i^R$ as:
\be
S_i^R(t)=\sum_{\tau=0}^\infty {\cal K}(\tau) \left[S_i(t)-\sum_{\tau'=1}^\tau 
\delta S_i(t-\tau')\right]=S_i(t) - \sum_{\tau'=1}^\infty \overline{{\cal 
K}}(\tau')
\delta S_i(t-\tau'),
\ee
where $\overline{{\cal K}}(\tau)$ is the (discrete) integral of ${\cal 
K}(\tau)$:
\be
\overline{{\cal K}}(\tau)=\sum_{\tau'=\tau}^\infty {\cal K}(\tau').
\ee
Note that from the normalization of ${\cal K}(\tau)$, one has
$\overline{{\cal K}}(0)=1$, independently of the specific shape of 
${\cal K}(\tau)$. This will turn out to be crucial in the following 
discussion.

For the exponential model, one finds
$\overline{{\cal K}}(\tau)=\alpha^\tau$. The limit $\alpha \to 1$
corresponds to the case where $S_i^R(t)$ is a constant, and therefore
Eq.~(\ref{retarded}) corresponds to an {\it additive} model. The other
limit $\alpha \to 0$ (infinitely small averaging time window) leads to
$S_i^R(t)\equiv S_i(t)$ and corresponds to a purely {\it
multiplicative} model. An value of $\alpha$ close to one, $\alpha=1-\epsilon$ 
corresponds
to an additive model on short time scales ($\ll T=\epsilon^{-1}$) and
to a multiplicative model for long time scales ($\gg T$) \cite{Book}. A 
formulation of this model in terms of product of random matrices is given in 
the Appendix.

In the following, we will assume that the relative difference between $S_i$ 
and $S_i^R$
is small. This is the case when:
\be
\eta = \sigma \sqrt{\sum_{\tau'=1}^\infty \overline{{\cal K}}^2(\tau')} \ll 1 
\ee
For the exponential model, this is equivalent to $\sigma\sqrt{T/2} \ll 1$. 
In other words, $\eta \ll 1$ means that the averaging takes place on a time 
scale 
such that the relative changes
of price are not too large. Typically, $T=50$ days, $\sigma=2 \%$ per 
square-root day, 
so that $\eta \sim 0.1$.

Let us calculate the correlation function ${\cal L}_{i}(\tau)$ to first order 
in $\eta \ll 1$. 
One finds, using $\delta x_i=\delta S_i/S_i$ and Eq.~(\ref{retarded}):
\be
\left\langle [\delta x_i(t+\tau)]^2 \  \delta x_i(t) \right\rangle = 
\sigma_i^2  
\left\langle \left(1 - 2\sum_{\tau'=1}^\infty \overline{{\cal K}}(\tau') 
\frac{\delta S_i(t+\tau-\tau')}{S_i(t+\tau)} \right) \frac{\delta 
S_i(t)}{S_i(t)} \right\rangle.
\ee
To first order in $\eta$, one can replace in the above expression 
$\delta S_i(t+\tau-\tau')/S_i(t+\tau)$ by $\sigma_i \epsilon(t+\tau-\tau')$ 
and 
$\delta S_i(t)/S_i(t)$ by $\sigma_i \epsilon(t)$. Since the $\epsilon$ are 
assumed to
be independent, of zero mean and of unit variance, one has
\be
\left\langle \epsilon(t+\tau-\tau')\, \epsilon(t) \right\rangle = 
\delta_{\tau,\tau'}.
\ee
Therefore:
\be
\left\langle [\delta x_i(t+\tau)]^2 \ \delta x_i(t) \right\rangle = 
- 2\sigma_i^4 \, \overline{{\cal K}}(\tau)
\ee
With the chosen normalization for ${\cal L}_{i}(\tau)$ (see Eq.~(\ref{C1})), 
we finally find:
\be
{\cal L}_{i}(\tau)= -2\overline{{\cal K}}(\tau)
\ee
A very important prediction of this model is therefore that ${\cal 
L}_{i}(\tau \to 0)= -2$. 
As shown in
Fig.~1, this is indeed rather well obeyed for individual stocks, with 
$\overline{{\cal K}}(\tau)$ given by a simple exponential. We have confirmed 
this finding 
by analyzing a set of 500 European stocks and 300 Japanese stocks, again in 
the period 1990-2000. 
The results are given in Figure 3
for the European stocks. We again find an exponential behavior with a time 
scale on the order of 40 days
and, more importantly, and initial values of ${\cal L}_{i}$ close to the 
retarded model
value $-2$ \cite{f2}. A similar graph was obtained for Japanese stocks as 
well, which is
interesting since this market did behave very differently both from the U.S. 
and European
markets during the investigated time period. For the Japanese market, the 
prediction 
${\cal L}(\tau \to 0)= -2$ not as good: an exponential fit of 
${\cal L}(\tau)$ gives and $A_S=1.5$ and $T_S=47$ days.  

We therefore conclude that
the leverage effect for stocks might not have a very deep economical 
significance, 
but can be assigned to a simple `retarded' effect, where
the change of prices are calibrated not on the instantaneous value of the 
price but on an
exponential moving average of the price. On the other hand, as Fig.~2 
reveals, the correlation
function for indices has a much larger value for $\tau=0$ and the above 
interpretation cannot
hold. We will discuss this in more details in the next section.

\begin{figure}
\psfrag{yaxis}[ct][ct]{\small ${\cal L}(\tau)$}
\psfrag{xaxis}[cb][cb]{\small $\tau$}
\psfrag{legend1}[l][l]{\small Empirical data}
\psfrag{legend2}[l][l]{\small Exponential fit}
\centerline{\epsfig{file=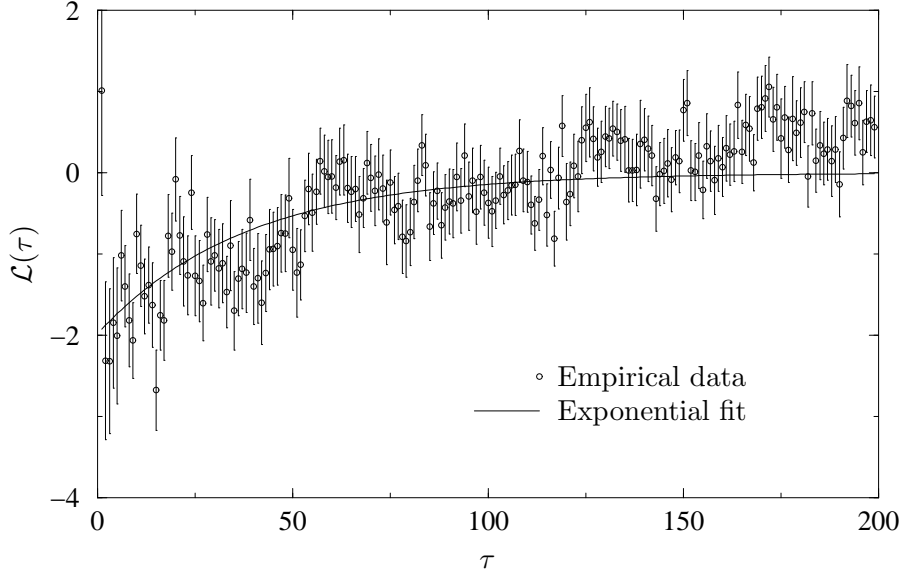,width=0.825\textwidth}}
\caption{Return-Volatility correlation for European
stocks. Data points are the empirical correlation averaged over 500 European 
stocks. 
The full line shows an exponential fit
(Eq.~(\protect\ref{exp_fit})) with $A_S=1.96$ and $T_S=38$ days for European 
stocks.
Note again that ${\cal L}$ for small $\tau$ is very close to the retarded 
model value $-2$.
}
\label{fig:fig_stocksbis}
\end{figure}

\section{A one factor leverage model}

Figure 2 shows that (i) the `leverage effect' for indices is much stronger 
than that 
appearing for individual stocks, but (ii) tends to decay to zero much faster 
with the lag
$\tau$. This is at first sight puzzling, since the stock index is, by 
definition, an average 
over stocks. So one can wonder why the strong leverage effect for the index 
does not show up 
in 
stocks and conversely why the long time scale seen in stocks disappears in 
the index. We want
to show in this section that these features can be rationalized by a simple 
one factor model,
where the `market factor' is responsible for the strong leverage effect.
Let us therefore write the stock price increments as:
\be
\delta S_i(t) = S_i^R(t) \, \left[\beta_i \phi(t) + 
\epsilon_i(t)\right],\label{modelf}
\ee
where $\phi(t)$ is the return factor common to all the stocks, $\beta_i$ are 
some time independent 
coefficients, and $\epsilon_i$ are the so-called 
idiosyncrasies, uncorrelated from stock to stock and from the common factor 
$\phi$.
The market index $I(t)$ is defined as a certain weighted average of the 
stocks:
\be
I(t) = \sum_{i=1}^N w_i S_i(t),
\ee
where $w_i$ are certain weights, of order $1/N$. From linearity, one finds 
the same relation 
between the `retarded' quantities:
\be
I^R(t)=\sum_{i=1}^N w_i S_i^R(t).
\ee
Neglecting terms of order $1/\sqrt{N}$, one finds, after summing 
Eq.~(\ref{modelf}) over all
$i$'s:
\be
\delta I(t)=\overline{\beta} I^R(t) \phi(t),
\ee
where $\overline{\beta}$ is the weighted sum of the $\beta$'s:
\be
\overline{\beta}=\frac{\sum_{i=1}^N w_i \beta_i S_i^R}{\sum_{i=1}^N w_i 
S_i^R},
\ee
that we will take as a constant. One can always define $\phi(t)$ such that 
this constant is unity, 
which is the choice we make in the following.
Now, we postulate that there exists an index specific leverage effect, 
resulting from an 
increase of activity when the market as a whole goes down, reflecting a panic 
like effect. 
Therefore we write:
\be
\left\langle \phi^2(t+\tau) \frac{\delta I(t)}{I(t)} \right\rangle \simeq 
\left\langle \phi^2(t+\tau) \phi(t) \right\rangle = - \Gamma(\tau)
\ee
In the following, we will indeed assume that $\Gamma(\tau)$ is small, as the
data suggests, and we neglect all mixed panic-retarded effects. 
To linear order in the correlations, one then
finds the following index leverage effect:
\be
\left\langle \left[\frac{\delta I(t+\tau)}{I(t+\tau)}\right]^2 
\frac{\delta I(t)}{I(t)} \right\rangle \simeq 
-\Gamma(\tau)-2\overline{{\cal K}}(\tau)
\left\langle \phi^2(t) \right\rangle^2.
\ee
Hence, the correctly normalized volatility-return correlation function reads:
\be
{\cal L}_{I}(\tau)=-2\overline{{\cal K}}(\tau)-\gamma(\tau)
\ee
with:
\be
\Gamma(\tau)= \gamma(\tau) \sigma_I^4 
\ee
where $\sigma_I^2 \equiv \sqrt{\langle \phi^2(t)\rangle}$ is the market 
volatility.
Therefore, one explicitly sees that the slowly decaying part
$\overline{{\cal K}}(\tau)$ should also appear in ${\cal L}_{I}(\tau)$. The
amplitude of this retarded correlation ($2\overline{{\cal K}}(0)=2$) is 
however
only $10\%$ of the observed correlation (${\cal L}_{I}(0)=18.$). We have 
fitted the
observed correlation for indices by a sum of two exponentials, with
only the parameters of the `fast' one left free, the slow one being
fixed by fitting individual stocks. The resulting fit (not shown) was
not significantly better (nor worse) than the single exponential fit of
Fig.~2.  Given the amount of noise in the data, it is difficult to
prove or disprove the presence of the slowly decaying correlation.
Nevertheless, we argue that it should be present for reasons of consistency 
between the index and its constituents.

Let us finally estimate the contribution of $\gamma(\tau)$ to the individual 
stock leverage
effect. A simple computation gives, to lowest order:
\be
{\cal L}_{i}(\tau)= -2\overline{{\cal K}}(\tau)-
\beta_i^3 \left(\frac{\sigma_I}{\sigma_i}\right)^4
\gamma(\tau).
\ee
Since the
market volatility $\sigma_I$ is a factor 3 smaller than the volatility of 
individual stocks $\sigma_i$ \cite{Cizeau}, the prefactor in front of 
$\gamma(\tau)$ is of order $1/100$. Therefore, even if $\gamma(\tau)$ is ten 
times larger than 
$\overline{{\cal K}}(\tau)$,
the influence of the market leverage effect on individual stocks is 
effectively suppressed due to relatively large 
ratio between the stock volatility and the market volatility. Again, due to 
the amount of 
noise in
 the data, it is difficult to confirm directly 
the presence of the $\gamma(\tau)$ contribution in ${\cal L}_{i}(\tau)$. 
However, since this contribution
is small and decays relatively fast, we believe that its role for individual 
stocks 
can safely be neglected.

\section{Conclusion}

In this paper, we have investigated quantitatively the so-called
leverage effect, which corresponds to a negative correlation between
future volatility and past returns. We have found that this correlation is
moderate and decays over a few months for individual stocks, and much
stronger but decaying much faster for stock indices. In the case of
individual stocks, we have found that the magnitude of this
correlation can be rationalized in terms of a retarded effect, which
assumes that the reference price used to set the scale for price
updates is {\it not} the instantaneous price but rather a moving
average of the price over the last few months. This interpretation, supported
by the data on U.S., European and Japanese stocks,
appears to us rather likely, and defines an interesting class of
stochastic processes, intermediate between purely additive (valid on short
time scales) and purely
multiplicative (relevant for long time scales), first advocated in 
\cite{Book}. 
For stock indices, however, this
interpretation breaks down and a specific market panic
phenomenon seems to be responsible for the enhanced observed negative
correlation between volatility and returns (and in turn to the strong
skews observed on index option smiles). Interestingly, this effect
appears to decay rather quickly, over a few days. As a simple one factor
model shows, the two effects
(retardation and panic) are not incompatible and should both be
present in individual stocks and stock indices. However the relative amplitude
of the retardation effect for indices and of the panic effect for
stocks are both too small to be unambiguously detected 
with only stock or index data.  Note that in both cases, 
the correlation between 
volatility and
returns appears to decay exponentially, in strong contrast to
volatility-volatility correlations which decay as a power-law. This
power-law behavior has recently lead to the construction of a
beautiful multifractal model \cite{muzy2}.  It would be interesting,
from a theoretical point of view, to generalize this model to account
for a scale invariant leverage effect. Work in this direction is underway.

\section*{Acknowledgments}

We thank M. Meyer and P. Cizeau for their commitment in an early stage of 
this work, 
J.P. Fouque for many interesting discussions about smiles and skews, and E. 
Bacry for
discussions about the multifractal models.

\section*{Appendix}

The exponential retarded model can be reformulated in terms of a product of 
random
$2 \times 2$ matrices. Define the vector $\vec V(t)$ as the set $S_R(t), \, 
S(t)$. 
One then has:
\be
\vec V(t+1) = {\cal M}(t) \vec V(t)
\ee
where ${\cal M}(t)$ is a random matrix whose elements are ${\cal 
M}_{11}(t)=\alpha$,
${\cal M}_{12}(t)=1-\alpha$, ${\cal M}_{21}(t)=\sigma \epsilon(t)$ and 
${\cal M}_{22}(t)=0$. Using standard results on products of random matrices, 
one
directly finds that the large time statistics of both components of $\vec V$ 
is log-normal.


\begin{thebibliography}{99}
\bibitem{f1} One can also consider the case where $\sigma_i$ 
is a random, time dependent quantity, to account for volatility clustering.
\bibitem{f2} We have noted however that the very first point $\tau=1$ day 
is systematically, and significantly, above the value $-2$. We have not found 
a simple
interpretation of this fact.
\end{thebibliography}
\end{document}